\documentclass[%
 reprint,
 amsmath,amssymb,
 aps,
]{revtex4-2}

\usepackage{xcolor}
\usepackage{graphicx}
\usepackage{dcolumn}
\usepackage{bm}
\usepackage{tikz}

\begin{document}

\preprint{APS/123-QED}

\title{Spin Inertia as a Driver of Chaotic and High-Speed Ferromagnetic Domain Walls}

\author{A.L. Bassant$^{1}$}
 \email{a.l.bassant@uu.nl.}
\author{Y.M.J. Ohlsen$^{1}$}
\author{M. Cherkasskii$^2$}
\author{P. B. He$^{3}$}
\author{R.A. Duine$^{1,4}$}
\affiliation{$^1$Institute for Theoretical Physics, Utrecht University, Princetonplein 5, 3584CC Utrecht, The Netherlands \\
$^2$Institute for Theoretical Solid State Physics, RWTH Aachen University, DE-52074 Aachen, Germany \\
$^3$School of Physics and Electronics, Hunan University, Changsha 410082, China \\
$^4$Department of Applied Physics, Eindhoven University of Technology, P.O. Box 513, 5600 MB Eindhoven, The Netherlands}

\date{\today}

\begin{abstract}

Ferromagnetic domain walls —\hspace{1px}transitional regions between magnetic domains\hspace{1px}— are an essential ingredient for racetrack memory, a device concept that promises to deliver faster and more compact memory storage compared to other non-volatile memory devices. 
Motivated by recent experiments that have found inertial effects in spin dynamics, we explore its consequences on domain wall motion.
We find that the inertial dynamics of the individual magnetic moments induce massive dynamics of the domain wall.
We investigate these massive dynamics driven by a magnetic field, spin-transfer torque, and spin-orbit torque.
We show that, in the absence of Gilbert damping, the domain wall dynamics become chaotic, resembling that of an electron in a two-dimensional crystal.
For finite damping, field-like driving of the inertial domain wall significantly increases its velocity compared to conventional massless dynamics, potentially enabling faster racetrack operations.
Additionally, in the limit of low driving, we observe that the domain wall width contracts due to the spin inertia of the ferromagnet.

\end{abstract}

\maketitle


\section{Introduction}

Domain walls are magnetic textures in between magnetic domains, and they are a key component in the development of future magnetic memory storage, such as magnetic racetrack memory \cite{RT:Parkin_Yang_2015,RT:Parkin_Hayashi_Thomas_2008,RT:Blasing_Khan_Filippou_Garg_Hameed_Castrillon_Parkin_2020}. 
To achieve optimal performance of racetrack memory, a thorough understanding of domain wall (DW) dynamics in ferromagnets is essential.
Their dynamics are often reduced to a small set of collective coordinates, typically the wall position and an internal angle describing the wall chirality \cite{Shibata_Tatara_Kohno_2011}. While the Landau-Lifshitz-Gilbert (LLG) equation with spin-transfer torque yields coupled first-order equations for these variables, an effective mass emerges once internal degrees of freedom are eliminated or when the wall must deform dynamically. In this sense, the DW mass is an effective parameter encoding the delayed response associated with internal modes, restoring torques, and pinning rather than a fundamental ingredient of the microscopic equations \cite{Tatara_Kohno_Shibata_2008}. In particular, massive DWs can originate from coupling to conducting electrons \cite{Mass:Hurst_Galitski_Heikkila_2020,Mass:Saitoh_Miyajima_Yamaoka_Tatara_2004}.
 Also, strong perpendicular anisotropy causes massive dynamics, where the internal angle of the DW acts as a slave variable to the DW position, in which case the mass is often discussed in terms of the Döring mass \cite{Mass:doring1948tragheit,Mass:Janak_1964}. 
 These examples of massive DW dynamics have also been shown to be chaotic.
Chaotic DW motion has been achieved by radio-frequency magnetic driving in the presence of pinning \cite{DWChaos:Pivano_Dolocan_2016,DWChaos:Suhl_Zhang_1987} or D\"oring mass \cite{DWChaos:Okuno_Hirata_Sakata_1995,DWChaos:Okuno_Sugitani_Hirata_1995,DWChoas:Kosinski_Sukiennicki_1992}. Additionally, chaotic DW motion can also be achieved using an alternating current \cite{DWChaos:Hermann_Nguenang_2013,DWChaos:Matsushita_Sasaki_Chawanya_2012}.

In planar nanostrips, the DW mass provides the channel underlying Walker breakdown, which acts as a dynamical speed limit for steady high-velocity DW propagation. However, it was shown that this limitation can be removed in cylindrical nanowires. Owing to rotational symmetry, a transverse DW can rotate about the wire axis without a demagnetizing-energy cost, eliminating a Walker-type critical threshold \cite{Yan2010,Yan_Andreas_Kakay_Garcia-Sanchez_Hertel_2011}. In this massless regime, the DW can propagate at velocities tracking the spin-drift scale up to very large current densities. Another experimental evidence for negligible DW inertia under short pulses was demonstrated in perpendicularly magnetized nanostripes using time-resolved imaging \cite{Vogel2012}. These results support the view that massless DW dynamics can be achieved either by symmetry or by suppressing internal deformation and ensuring rapid relaxation of internal modes. 

A distinct notion of inertia has also been introduced at the level of the magnetization dynamics itself. Recent experimental discoveries have been pushing magnetic dynamics into high-frequency ranges where conventional approaches cannot be applied~\cite{KNeeraj,ADe,YLi,VUnikandanunni,YLi}. Already at sub-terahertz frequencies, the LLG equation becomes insufficient because the alignment between the magnetization and the angular momentum is no longer instantaneous. The resulting finite response time induces a temporal lag that is captured by augmenting the LLG equation with an inertial term proportional to the second time derivative of the magnetization equation~\cite{Suhl1998,wegrowe2000thermokinetic,Ciornei2011,Wegrowe2012,giordano2020derivation}. The resulting equation is called the inertial Landau-Lifshitz-Gilbert (ILLG) equation. 

Relativistic quantum derivations corroborate the ILLG equation by recovering, from first principles, the same magnetization second-time-derivative term~\cite{Fahnle2011,Bhattacharjee2012,Mondal2017Nutation,Mondal2018JPCM}. Ab initio calculations show that both Gilbert damping and inertial terms emerge self-consistently from the electronic structure, with spin--orbit coupling providing the common microscopic origin of dissipative and inertial magnetization dynamics~\cite{Bhattacharjee2012,RMondal_JPCM,RMondal}. Another approach treats the localized spins as coupled to an environmental bath. This approach shows that a non-Markovian memory kernel in the bath response generates an effective inertial term in the magnetization equation of motion~\cite{MAQuarenta,JAnders}. Nonadiabatic coupling to environmental degrees of freedom provides an alternative route to derive the inertial extension of the LLG equation~\cite{TKikuchi}. The magnetic inertia introduces an additional nutational resonance, typically in the sub-THz range, distinct from the conventional precessional resonance at GHz frequencies~\cite{EOlive_APL,EOlive_JAP,SVTitov,MCherkasskii_PRB102,MCherkasskii_PRB106}. The clearest experimental signatures are high-frequency nutation responses observed in NiFe and CoFeB~\cite{KNeeraj,ADe,YLi} and in Co~\cite{VUnikandanunni,YLi} films.

Magnetic inertia has also been investigated in heavy-metal/ferromagnet heterostructures that generate spin--orbit torques. A framework for precessional and nutational auto-oscillations in inertial ferromagnets has been developed~\cite{RRodriguez}, and auto-oscillations have been studied in both ferromagnets~\cite{RRodriguez} and antiferromagnets~\cite{PBHe_PRB108}. The influence of inertia on self-oscillation has further been examined in spin--orbit-torque-driven tripartite antiferromagnets with $120^{\circ}$ rotational symmetry~\cite{PBHe_PRB110}. While spin inertia has thus been explored in a range of applications, its potential impact on domain wall dynamics remains unexplored.

In this paper, we theoretically show that spin inertia causes a ferromagnetic DW to be effectively massive. In contrast to previous approaches, we introduce the DW mass through inertial spin dynamics and study DW motion driven by magnetic fields, spin-transfer torque (STT), and spin-orbit torque (SOT). This approach allows us to analyze massive DW dynamics under both field-like and damping-like torque components. We find that the resulting torque balance can lead to nontrivial behavior, including chaotic trajectories and substantially enhanced DW velocities.

In the absence of damping, the DW equations of motion map onto those of a charged particle moving in a two-dimensional potential that is periodic in one direction and subject to a perpendicular magnetic field. If the potential is periodic in both directions, the mapping becomes analogous to an electron in a two-dimensional crystal under a perpendicular magnetic field. This system displays chaotic dynamics and, upon quantization, yields the Hofstadter butterfly spectrum \cite{HofstadterButterfly}. Within our classical model, we demonstrate chaos by evaluating the Lyapunov exponent.

We also consider damped DW dynamics. In this case, dissipation regularizes the motion and suppresses chaos. For specific driving strengths, the DW velocity attains a maximum that is substantially larger than in the non-inertial case. Moreover, in the low-driving limit we find that spin inertia reduces the DW width.

These features may be relevant for applications based on ferromagnetic DWs. In particular, racetrack memory could benefit from operating regimes in which inertial effects enhance DW velocities. In addition, chaotic massive DW motion may serve as a controllable experimental analog of charged-particle dynamics in two-dimensional periodic potentials, providing insight into transport in such systems.

The rest of this paper is organized as follows. Section \ref{sec:Model} introduces the model and the equations of motion. In Section \ref{sec:Chaotic dynamics}, we derive the dissipationless Hamiltonian and prove that its dynamics are chaotic by calculating the Lyapunov exponent. Thereafter, in Section \ref{sec:Domain wall velocities}, we take into account dissipation and compute domain wall velocities for different types of driving and inertia. In the low driving limit we show that the DW width decreases, which is shown in Section \ref{sec:DWwidth}. Finally, we discuss the results in Section \ref{sec:Discussion and Conclusion}.

\section{Model}\label{sec:Model}

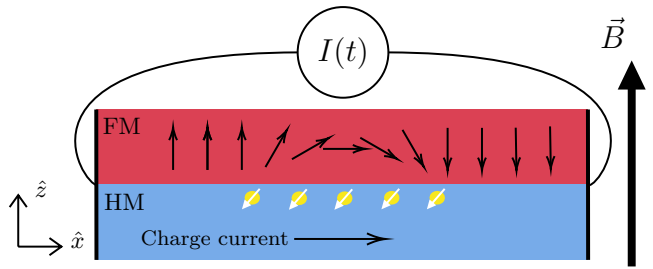
\begin{figure}
    \centering

\tikzset{every picture/.style={line width=0.75pt}} 

\begin{tikzpicture}[x=0.65pt,y=0.65pt,yscale=-1,xscale=1]

\draw  [draw opacity=0][fill={rgb, 255:red, 74; green, 144; blue, 226 }  ,fill opacity=0.75 ] (188.52,178.85) -- (474.64,178.85) -- (474.64,222.85) -- (188.52,222.85) -- cycle ;
\draw  [draw opacity=0][fill={rgb, 255:red, 208; green, 2; blue, 27 }  ,fill opacity=0.75 ] (188.41,135.25) -- (474.64,135.25) -- (474.64,178.85) -- (188.41,178.85) -- cycle ;
\draw    (253.12,170.45) -- (253.12,147.45) ;
\draw [shift={(253.12,145.45)}, rotate = 90] [color={rgb, 255:red, 0; green, 0; blue, 0 }  ][line width=0.75]    (7.65,-2.3) .. controls (4.86,-0.97) and (2.31,-0.21) .. (0,0) .. controls (2.31,0.21) and (4.86,0.98) .. (7.65,2.3)   ;
\draw    (273.12,170.45) -- (273.12,150.95) -- (273.12,147.45) ;
\draw [shift={(273.12,145.45)}, rotate = 90] [color={rgb, 255:red, 0; green, 0; blue, 0 }  ][line width=0.75]    (7.65,-2.3) .. controls (4.86,-0.97) and (2.31,-0.21) .. (0,0) .. controls (2.31,0.21) and (4.86,0.98) .. (7.65,2.3)   ;
\draw    (286.87,168.78) -- (298.37,148.86) ;
\draw [shift={(299.37,147.12)}, rotate = 120] [color={rgb, 255:red, 0; green, 0; blue, 0 }  ][line width=0.75]    (7.65,-2.3) .. controls (4.86,-0.97) and (2.31,-0.21) .. (0,0) .. controls (2.31,0.21) and (4.86,0.98) .. (7.65,2.3)   ;
\draw    (302.29,164.2) -- (322.21,152.7) ;
\draw [shift={(323.94,151.7)}, rotate = 150] [color={rgb, 255:red, 0; green, 0; blue, 0 }  ][line width=0.75]    (7.65,-2.3) .. controls (4.86,-0.97) and (2.31,-0.21) .. (0,0) .. controls (2.31,0.21) and (4.86,0.98) .. (7.65,2.3)   ;
\draw    (320.37,158.2) -- (343.37,158.2) ;
\draw [shift={(345.37,158.2)}, rotate = 180] [color={rgb, 255:red, 0; green, 0; blue, 0 }  ][line width=0.75]    (7.65,-2.3) .. controls (4.86,-0.97) and (2.31,-0.21) .. (0,0) .. controls (2.31,0.21) and (4.86,0.98) .. (7.65,2.3)   ;
\draw    (342.04,151.7) -- (358.93,161.45) -- (361.96,163.2) ;
\draw [shift={(363.69,164.2)}, rotate = 210] [color={rgb, 255:red, 0; green, 0; blue, 0 }  ][line width=0.75]    (7.65,-2.3) .. controls (4.86,-0.97) and (2.31,-0.21) .. (0,0) .. controls (2.31,0.21) and (4.86,0.98) .. (7.65,2.3)   ;
\draw    (366.62,147.12) -- (378.12,167.04) ;
\draw [shift={(379.12,168.78)}, rotate = 240] [color={rgb, 255:red, 0; green, 0; blue, 0 }  ][line width=0.75]    (7.65,-2.3) .. controls (4.86,-0.97) and (2.31,-0.21) .. (0,0) .. controls (2.31,0.21) and (4.86,0.98) .. (7.65,2.3)   ;
\draw    (392.87,145.95) -- (392.87,168.95) ;
\draw [shift={(392.87,170.95)}, rotate = 270] [color={rgb, 255:red, 0; green, 0; blue, 0 }  ][line width=0.75]    (7.65,-2.3) .. controls (4.86,-0.97) and (2.31,-0.21) .. (0,0) .. controls (2.31,0.21) and (4.86,0.98) .. (7.65,2.3)   ;
\draw    (412.87,145.95) -- (412.87,168.95) ;
\draw [shift={(412.87,170.95)}, rotate = 270] [color={rgb, 255:red, 0; green, 0; blue, 0 }  ][line width=0.75]    (7.65,-2.3) .. controls (4.86,-0.97) and (2.31,-0.21) .. (0,0) .. controls (2.31,0.21) and (4.86,0.98) .. (7.65,2.3)   ;
\draw  (143,215.12) -- (168.18,215.12)(143,185.1) -- (143,215.12) -- cycle (161.18,210.12) -- (168.18,215.12) -- (161.18,220.12) (138,192.1) -- (143,185.1) -- (148,192.1)  ;
\draw    (253.12,170.45) -- (253.12,150.95) -- (253.12,147.45) ;
\draw [shift={(253.12,145.45)}, rotate = 90] [color={rgb, 255:red, 0; green, 0; blue, 0 }  ][line width=0.75]    (7.65,-2.3) .. controls (4.86,-0.97) and (2.31,-0.21) .. (0,0) .. controls (2.31,0.21) and (4.86,0.98) .. (7.65,2.3)   ;
\draw    (432.5,145.85) -- (432.91,168.85) ;
\draw [shift={(432.94,170.85)}, rotate = 268.99] [color={rgb, 255:red, 0; green, 0; blue, 0 }  ][line width=0.75]    (7.65,-2.3) .. controls (4.86,-0.97) and (2.31,-0.21) .. (0,0) .. controls (2.31,0.21) and (4.86,0.98) .. (7.65,2.3)   ;
\draw    (452.49,145.1) -- (452.9,168.1) ;
\draw [shift={(452.93,170.1)}, rotate = 268.99] [color={rgb, 255:red, 0; green, 0; blue, 0 }  ][line width=0.75]    (7.65,-2.3) .. controls (4.86,-0.97) and (2.31,-0.21) .. (0,0) .. controls (2.31,0.21) and (4.86,0.98) .. (7.65,2.3)   ;
\draw   (305.61,102) .. controls (305.61,87.42) and (317.43,75.6) .. (332.01,75.6) .. controls (346.59,75.6) and (358.41,87.42) .. (358.41,102) .. controls (358.41,116.58) and (346.59,128.4) .. (332.01,128.4) .. controls (317.43,128.4) and (305.61,116.58) .. (305.61,102) -- cycle ;
\draw    (474.81,179.44) .. controls (483.67,179.67) and (532.67,102.67) .. (358.41,102) ;
\draw [line width=1.5]    (188.41,135.25) -- (188.52,222.85) ;
\draw [line width=1.5]    (474.53,135.25) -- (474.64,222.85) ;
\draw  [draw opacity=0][fill={rgb, 255:red, 248; green, 231; blue, 28 }  ,fill opacity=1 ] (329.47,184.49) .. controls (330.98,182.75) and (333.79,182.43) .. (335.74,183.78) .. controls (337.7,185.12) and (338.06,187.62) .. (336.55,189.36) .. controls (335.03,191.1) and (332.22,191.42) .. (330.27,190.07) .. controls (328.31,188.73) and (327.96,186.23) .. (329.47,184.49) -- cycle ;
\draw [color={rgb, 255:red, 255; green, 255; blue, 255 }  ,draw opacity=1 ][fill={rgb, 255:red, 248; green, 231; blue, 28 }  ,fill opacity=1 ]   (337.35,181.74) -- (328.95,191.4) ;
\draw [shift={(326.98,193.66)}, rotate = 310.99] [fill={rgb, 255:red, 255; green, 255; blue, 255 }  ,fill opacity=1 ][line width=0.08]  [draw opacity=0] (5.36,-2.57) -- (0,0) -- (5.36,2.57) -- cycle    ;
\draw  [draw opacity=0][fill={rgb, 255:red, 248; green, 231; blue, 28 }  ,fill opacity=1 ] (303.01,184.49) .. controls (304.52,182.75) and (307.33,182.43) .. (309.28,183.78) .. controls (311.24,185.12) and (311.6,187.62) .. (310.09,189.36) .. controls (308.57,191.1) and (305.76,191.42) .. (303.81,190.07) .. controls (301.85,188.73) and (301.5,186.23) .. (303.01,184.49) -- cycle ;
\draw [color={rgb, 255:red, 255; green, 255; blue, 255 }  ,draw opacity=1 ][fill={rgb, 255:red, 248; green, 231; blue, 28 }  ,fill opacity=1 ]   (310.89,181.74) -- (302.49,191.4) ;
\draw [shift={(300.52,193.66)}, rotate = 310.99] [fill={rgb, 255:red, 255; green, 255; blue, 255 }  ,fill opacity=1 ][line width=0.08]  [draw opacity=0] (5.36,-2.57) -- (0,0) -- (5.36,2.57) -- cycle    ;
\draw  [draw opacity=0][fill={rgb, 255:red, 248; green, 231; blue, 28 }  ,fill opacity=1 ] (275.65,184.23) .. controls (277.17,182.49) and (279.98,182.17) .. (281.93,183.51) .. controls (283.89,184.86) and (284.25,187.36) .. (282.73,189.09) .. controls (281.22,190.83) and (278.41,191.15) .. (276.46,189.81) .. controls (274.5,188.46) and (274.14,185.96) .. (275.65,184.23) -- cycle ;
\draw [color={rgb, 255:red, 255; green, 255; blue, 255 }  ,draw opacity=1 ][fill={rgb, 255:red, 248; green, 231; blue, 28 }  ,fill opacity=1 ]   (283.53,181.47) -- (275.14,191.13) ;
\draw [shift={(273.17,193.4)}, rotate = 310.99] [fill={rgb, 255:red, 255; green, 255; blue, 255 }  ,fill opacity=1 ][line width=0.08]  [draw opacity=0] (5.36,-2.57) -- (0,0) -- (5.36,2.57) -- cycle    ;
\draw  [draw opacity=0][fill={rgb, 255:red, 248; green, 231; blue, 28 }  ,fill opacity=1 ] (383.28,184.23) .. controls (384.79,182.49) and (387.6,182.17) .. (389.55,183.51) .. controls (391.51,184.86) and (391.87,187.36) .. (390.36,189.09) .. controls (388.85,190.83) and (386.04,191.15) .. (384.08,189.81) .. controls (382.13,188.46) and (381.77,185.96) .. (383.28,184.23) -- cycle ;
\draw [color={rgb, 255:red, 255; green, 255; blue, 255 }  ,draw opacity=1 ][fill={rgb, 255:red, 248; green, 231; blue, 28 }  ,fill opacity=1 ]   (391.16,181.47) -- (382.76,191.13) ;
\draw [shift={(380.79,193.4)}, rotate = 310.99] [fill={rgb, 255:red, 255; green, 255; blue, 255 }  ,fill opacity=1 ][line width=0.08]  [draw opacity=0] (5.36,-2.57) -- (0,0) -- (5.36,2.57) -- cycle    ;
\draw  [draw opacity=0][fill={rgb, 255:red, 248; green, 231; blue, 28 }  ,fill opacity=1 ] (356.82,184.23) .. controls (358.33,182.49) and (361.14,182.17) .. (363.09,183.51) .. controls (365.05,184.86) and (365.41,187.36) .. (363.9,189.09) .. controls (362.39,190.83) and (359.58,191.15) .. (357.62,189.81) .. controls (355.67,188.46) and (355.31,185.96) .. (356.82,184.23) -- cycle ;
\draw [color={rgb, 255:red, 255; green, 255; blue, 255 }  ,draw opacity=1 ][fill={rgb, 255:red, 248; green, 231; blue, 28 }  ,fill opacity=1 ]   (364.7,181.47) -- (356.3,191.13) ;
\draw [shift={(354.33,193.4)}, rotate = 310.99] [fill={rgb, 255:red, 255; green, 255; blue, 255 }  ,fill opacity=1 ][line width=0.08]  [draw opacity=0] (5.36,-2.57) -- (0,0) -- (5.36,2.57) -- cycle    ;
\draw    (303.59,210.65) -- (354.86,210.67) ;
\draw [shift={(356.86,210.67)}, rotate = 180.02] [color={rgb, 255:red, 0; green, 0; blue, 0 }  ][line width=0.75]    (10.93,-3.29) .. controls (6.95,-1.4) and (3.31,-0.3) .. (0,0) .. controls (3.31,0.3) and (6.95,1.4) .. (10.93,3.29)   ;
\draw [line width=3]    (500,226.67) -- (500,112.67) ;
\draw [shift={(500,106.67)}, rotate = 90] [fill={rgb, 255:red, 0; green, 0; blue, 0 }  ][line width=0.08]  [draw opacity=0] (16.97,-8.15) -- (0,0) -- (16.97,8.15) -- cycle    ;
\draw    (233.12,170.45) -- (233.12,150.95) -- (233.12,147.45) ;
\draw [shift={(233.12,145.45)}, rotate = 90] [color={rgb, 255:red, 0; green, 0; blue, 0 }  ][line width=0.75]    (7.65,-2.3) .. controls (4.86,-0.97) and (2.31,-0.21) .. (0,0) .. controls (2.31,0.21) and (4.86,0.98) .. (7.65,2.3)   ;
\draw    (189.21,179.44) .. controls (180.35,179.67) and (131.35,102.67) .. (305.61,102) ;

\draw (151.01,174.68) node [anchor=north west][inner sep=0.75pt]    {$\hat{z}$};
\draw (171.91,204.8) node [anchor=north west][inner sep=0.75pt]    {$\hat{x}$};
\draw (315.41,91.4) node [anchor=north west][inner sep=0.75pt]  [font=\large]  {$I(t)$};
\draw (190.41,138.25) node [anchor=north west][inner sep=0.75pt]   [align=left] {FM};
\draw (191.21,182.44) node [anchor=north west][inner sep=0.75pt]   [align=left] {HM};
\draw (212.66,203.94) node [anchor=north west][inner sep=0.75pt]  [font=\footnotesize] [align=left] {Charge current};
\draw (480,80.07) node [anchor=north west][inner sep=0.75pt]  [font=\large]  {$\vec{B}$};

\end{tikzpicture}

    \caption{The setup is given by a conducting ferromagnetic (FM) and heavy metal (HM) layer. The ferromagnet has a domain wall texture, which is driven by the spin transfer torque, spin orbit torque, and an external magnetic field $\vec B$.}
    \label{fig:setup}
\end{figure}

In this section, we derive the equations of motion for a DW in a ferromagnetic wire with large spin inertia.
The ferromagnetic wire is influenced by an external magnetic field, STT, and SOT as seen in Fig. \ref{fig:setup}.
We start with the ILLG equation, which follows from the effective ferromagnetic action.
Then we use the DW ansatz to derive its action, which yields the equations of motion for the DW's position and internal angle. 

The dynamics of the ferromagnet's magnetization, $\boldsymbol{\Omega}$, are governed by the LLG equation with STT and SOT.
Inertial dynamics are known to enter as follows \cite{Ciornei_Rubi_Wegrowe_2011},

\begin{align}
    \label{iLLG equation}
    \left(\partial_t+v_s\partial_x\right)\boldsymbol{\Omega} =& - \gamma \boldsymbol{\Omega} \times \boldsymbol{H}_{eff}-\sigma \tau( \boldsymbol{\Omega}\times \hat z)\nonumber\\
    &-\tau\boldsymbol{\Omega}\times(\boldsymbol{\Omega}\times \hat z)+\eta\boldsymbol{\Omega}\times \partial_t^2\boldsymbol{\Omega} \nonumber\\
    & + \alpha \boldsymbol{\Omega}\times\left(\partial_t+\frac{\beta}{\alpha}v_s\partial_x\right)\boldsymbol{\Omega},
\end{align}

\noindent where $\boldsymbol{\Omega}$ is the unit vector of magnetization, $\gamma$ is the gyromagnetic ratio, $v_s= gP\mu_{\mathrm{B}}j^{FM} / (2|e|M_s)$ describes the strength of the STT, $\tau=(\gamma/M_s) j_{s}^{SHE}/t_F$ denotes the strength of the SOT, $\alpha$ is the Gilbert damping, $\beta$ denotes the relative strength of the dissipative STT to the reactive one, and $\sigma$ is the ratio of the field-like to damping-like SOT.
The inertial dynamics are governed by $\eta$, which is called the inertial relaxation time.
The STT depends on the Land\'e factor $g$, current polarization $P$, Bohr magneton $\mu_{\mathrm{B}}$, charge current density in the ferromagnetic layer $j^{FM}$, and elementary charge $e$.
The SOT is caused by the spin current density generated by the spin Hall effect in the heavy metal $j_{s}^{HM}=(\hbar /2e)\eta_{eff}\theta_{sh}j^{HM}$ with $\eta_{eff}$ the spin-injection efficiency, $\theta_{sh}$ the spin Hall angle, and $j^{HM}$ the charge current density in the heavy metal layer~\cite{Ando_2008}.
Additionally, the parameter $\tau$ is inversely proportional to the thickness of the ferromagnet $t_F$.
The conjunction of STT and SOT has been investigated before in Ref. \cite{STTandSOT:Seo_Kim_Ryu_Lee_Lee_2012}.
The effective field is given by $\boldsymbol{H}_{eff}=\delta E/(M_s\delta\boldsymbol{\Omega})$ where $E$ is the energy functional,

\begin{align}
\label{eq:E}
    E[\boldsymbol{\Omega}] =  \int \frac{d \boldsymbol{x}}{a^3}\bigg\{ & \frac{J_s}{2}\left(\nabla \boldsymbol{\Omega}\right)^2-M_s B\Omega_z +\frac{K_\perp}{2}\Omega_y^2  \nonumber\\
    &- \frac{K_z}{2}\Omega_z^2\bigg\}, \nonumber \\
    =  \int \frac{d \boldsymbol{x}}{a^3}\bigg\{&\frac{J_s}{2}\left(\left(\nabla \theta\right)^2+\sin^2\theta\left(\nabla \phi\right)^2\right)-M_s B\cos\theta \nonumber\\
    &+\frac{K_\perp}{2}\sin^2\theta\sin^2\phi - \frac{K_z}{2}\cos^2\theta\bigg\}.
\end{align}

\noindent Here the spin stiffness is given by $J_s$, the lattice constant is given by $a$, the anisotropies are given by $K_\perp$ and $K_z$, and the external magnetic field in the $z$-direction is given by $B$.
The energy functional is also expressed in spherical coordinates, which are related to the magnetization by $\boldsymbol{\Omega}=(\cos(\phi)\sin(\theta),\sin(\phi)\sin(\theta),\cos(\theta))$.
The spherical coordinates are used to describe the effective action for the magnetization, which is given by,

\begin{align}
\label{eq:Am}
     \mathcal{A}_{FM}[\theta, \phi]=& \int dt\int\frac{d\boldsymbol{x}}{a^3}\Big\{\hbar (\cos(\theta) - 1)(\partial_t+v_s\partial_x)\phi \\
      + \frac{\hbar\eta}{2}&\left[\dot{\theta}^2   + \dot{\phi}^2  \sin^2(\theta)  \right]+\hbar \sigma \tau\cos(\theta)-\mathcal{E}[\theta,\phi]\Big\}.\nonumber
\end{align}

\noindent The action describes only conserving dynamics, and non-conserving dynamics are introduced through the Rayleigh dissipation functional,

\begin{align}\label{eq:Rm}
    \mathcal{R}_{FM}[\theta,\phi]&=\int dt \int\frac{d\boldsymbol{x}}{a^3} \Big\{\frac{\hbar \alpha}{2}\left[\left(\partial_t-\frac{\beta}{\alpha}v_s\partial_x\right) \theta\right]^2 \\
    +\frac{\hbar \alpha}{2}&\sin^2(\theta)\left[\left(\partial_t-\frac{\beta}{\alpha}v_s\partial_x\right) \phi\right]^2+\hbar \tau\dot\phi\sin^2(\theta)\Big\}. \nonumber
\end{align}

\noindent The ILLG equation with STT and SOT is recovered by computing $\delta\mathcal{A}_{FM}[\boldsymbol{\Omega}]/\delta\boldsymbol{\Omega}(\boldsymbol{x},t) = \delta \mathcal{R}_{FM}[\boldsymbol{\Omega}]/\delta\partial_t\boldsymbol{\Omega}(\boldsymbol{x}, t)$.

From the previous action and dissipation functional, we compute the action and dissipation functional for a DW.
This is achieved by substituting the DW ansatz and integrating over spatial variables.
We assume that the wire cross section is sufficiently small that the dynamics of the magnetization is effectively homogeneous in the $y$ and $z$ direction.
The DW ansatz is given by $\phi(x,t)=\phi_0(t)$ and $\theta(x,t)=2\arctan(\exp((x-r(t))/\lambda)$ with $\lambda=\sqrt{J/K_z}$.
In Section \ref{sec:DWwidth}, we show that for small driving, the DW width is slightly modified.
The variables of the DW angle and position are given by $\phi_0(t)$, and $r(t)$, respectively.
After substituting the ansatz and integrating over the spatial variables, we find that

\begin{align}
\label{eq:A}
     \mathcal{A}[r,\phi_0] =&N\int dt\Big\{\left(M_s B+\hbar\sigma \tau-\hbar \dot\phi_0(t)\right)\frac{r(t)}{\lambda}\nonumber\\
    & -\hbar \frac{v_s}{\lambda}\phi_0(t) -\frac{K_\perp}{2}\sin^2(\phi_0(t)) \nonumber \\
    &+\frac{\hbar\eta}{2}\left[\frac{\dot r^2(t)}{\lambda^2}+\dot\phi_0^2(t)\right]\Big\},
\end{align}

\noindent with the Rayleigh's dissipation functional,

 \begin{align}\label{eq:R}
    \mathcal{R}[r,\phi_0]=&N\int dt \Bigg\{\frac{\hbar \alpha}{2\lambda^2}\left(\frac{\beta}{\alpha}v_s-\dot r(t)\right)^2\nonumber \\
    &+\frac{\hbar \alpha}{2}\dot\phi_0^2(t)+\hbar\tau\dot\phi_0(t)\Bigg\}. 
\end{align}

\noindent Here $N=2\lambda S/a^3$ with $S$ the cross section of the ferromagnetic wire.
The newly derived action and dissipation functional govern the equations of motion of the DW variables, which are found by computing,

\begin{align}
    \frac{\delta\mathcal{A}[r,\phi_0]}{\delta r(t)} =& \frac{\delta \mathcal{R}[r,\phi_0]}{\delta\dot r(t)}, & \frac{\delta\mathcal{A}[r,\phi_0]}{\delta \phi_0(t)} =& \frac{\delta \mathcal{R}[r,\phi_0]}{\delta\dot \phi_0(t)}.
\end{align}

\noindent The DW variables are made unitless through the definition,

\begin{equation}
\label{Eq: dimensionless time}
    \dot{\widetilde{r}} \equiv \frac{d\widetilde{r}}{dt} = \frac{\dot{r}}{K_\perp\lambda/(2\hbar)},
\end{equation}

\noindent which is composed of dimensionless position, $\widetilde{r} = r/\lambda$, differentiated with respect to dimensionless time, $\widetilde{t} = t/(2\hbar/K_\perp)$.
The dimensionless variables are governed by the following equations of motion,

\begin{subequations}
\begin{align} \label{eq:eomr}
      \frac{\eta K_\perp}{2\hbar} \ddot{\widetilde{r}}(t) =& -\alpha\dot{\widetilde{r}}(t) -\dot{\phi}_0(t) + \frac{2(M_s B+\hbar \sigma \tau)}{K_\perp} \\ \label{eq:eomphi}
      & + \frac{2\hbar\beta v_s}{K_\perp\lambda}, \nonumber\\
      \frac{\eta K_\perp}{2\hbar}\ddot{\phi}_0(t) =& -\alpha\dot{\phi}_0(t) +\dot{\widetilde{r}}(t)-\sin(2\phi_0(t)) \\
      & - \frac{2\hbar v_s}{K_\perp\lambda}-\frac{2\hbar \tau}{K_\perp}. \nonumber
\end{align}
\end{subequations}

\noindent The DW variables are coupled, and current and field drive the DW position and angle.
The position is directly affected by field-like torques, which are caused by the magnetic field ($B$), and field-like effects from STT and SOT ($\beta v_s$, and $\sigma \tau$).
The angle of the DW is directly driven by damping-like torques from STT and SOT ($v_s$, and $\tau$).
The appearance of second time-derivatives in Eqs. (\ref{eq:eomr}-\ref{eq:eomphi}) directly stems from the spin inertia.
This term will act like a mass, which we show in the following sections.
First, we consider the case without damping, where chaotic dynamics dominate.
Afterwards, we look at nonvanishing damping, which regularizes the DW motion.
In particular, we will see that there exist specific values of driving where the DW exceeds non-inertial DW velocities.

\section{Chaotic dynamics}\label{sec:Chaotic dynamics}

In this section, we investigate the chaos that arises from inertia when damping is vanishingly small.
First, we derive the Hamiltonian associated with the equations of motion, which allows us to derive the canonical momentum of $\tilde r$ and $\phi_0$.
This leads to a set of four first order differential equations, from which we compute the Lyapunov exponent —a measure for chaos.

The equations of motion given in Eqs. (\ref{eq:eomr}-\ref{eq:eomphi}) are recast into the following form

\begin{equation}
    \label{Eq: Eq of motion vector notation}
    m\ddot{\boldsymbol{R}}(t) = -\alpha\dot{\boldsymbol{R}}(t) - \dot{\boldsymbol{R}}(t) \times \hat z_R -\frac{\partial V(\widetilde{r}(t), \phi_0(t))}{\partial\boldsymbol{R}},
\end{equation}

\noindent where $m = \eta K_\perp/2\hbar$, and

\begin{align}
    V(\widetilde{r}, \phi_0) =& -\left(\frac{2(M_s B +\hbar \sigma \tau)}{K_\perp}+ \frac{2\hbar\beta v_s}{\lambda K_\perp}\right)\widetilde{r}(t)  \nonumber\\
    &+ \left(\frac{2\hbar v_s}{\lambda K_\perp }+\frac{2\hbar\tau}{K_\perp}\right)\phi_0(t) +\sin^2(\phi_0(t)),  \nonumber\\
    =& -\mu_r\tilde r+\mu_\phi\phi_0 +\sin^2\phi_0,
\end{align}

\noindent and
 
\begin{equation}
\boldsymbol{R} = 
    \begin{pmatrix}
        \widetilde{r} \\
        \phi_0 \\
        0
    \end{pmatrix},
    \quad \hat z_R =
    \begin{pmatrix}
        0\\
        0\\
        1
    \end{pmatrix}.
\end{equation}

\noindent The potential $V(\widetilde{r}, \phi_0)$ contains driving terms, which we define as field-like and damping-like driving, $\mu_r$, and $\mu_\phi$, respectively.
The motion that is described by Eq. (\ref{Eq: Eq of motion vector notation}) is equivalent to that of a charged particle in a periodic potential, $V(\widetilde{r}, \phi_0)$, and perpendicular magnetic field, $\hat z_R$, with linear friction, $\alpha$.
If the periodic potential were periodic in both directions, then these equations would be equivalent to an electron in a 2-dimensional lattice with a perpendicular magnetic field.
It is known that the motion of these electrons is chaotic in the classical limit \cite{Chaos:Petschel_Geisel_1993,Chaos:Wagenhuber_Geisel_Niebauer_Obermair_1992}.

In the following analysis, we show that the solutions of Eq. (\ref{Eq: Eq of motion vector notation}) are chaotic when there is no friction.
We already established that this equation describes the motion of a charged particle in a potential, which allows us to construct the corresponding Hamiltonian,

\begin{align}\label{eq:chaoticH}
    H=\frac{(\boldsymbol{P}-\boldsymbol{A})^2}{2m}+V(\tilde r,\phi_0).
\end{align}

\noindent Herein, $\boldsymbol{P}=m\dot{\boldsymbol{R}}+\boldsymbol{A}=(p_r,p_\phi)$ and $\boldsymbol{A}=(0,-\tilde r,0)$ such that $\hat z_R=-\nabla\times \boldsymbol{A}$.
From Hamilton's equation, we find four first order differential equations for $\tilde r$, $\phi_0$, and their conjugate momenta $p_r$, $p_\phi$.

\begin{subequations}

\begin{align}
    \dot{\tilde r}=&\frac{1}{m}p_r, & \dot p_r=&\mu_r-\frac{1}{m}(p_\phi+\tilde r), \label{eq:Ch1} \\
    \dot{\phi}_0=&\frac{1}{m}(p_\phi+\tilde r), & \dot p_\phi=&-\mu_\phi-\sin 2\phi_0. \label{eq:Ch2}
\end{align}    
\end{subequations}

\begin{figure}
    \centering
    \includegraphics[width=\linewidth]{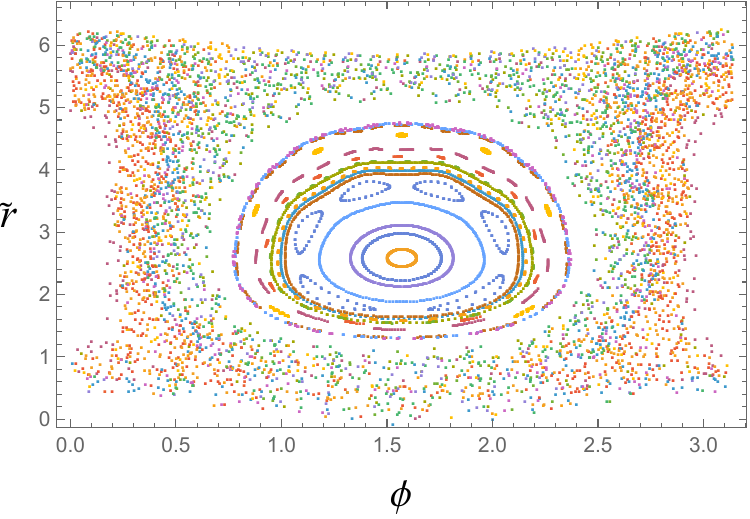}
    \caption{The Poincar\'e section shows 33 trajectories, each represented by a different colour. Every trajectory is generated by Eqs. (\ref{eq:Ch1}-\ref{eq:Ch1}) using different initial conditions with $H\approx 1$ (Eq. \ref{eq:chaoticH}). Each time the four-dimensional trajectory passes through a hyperplane defined by $\dot{\tilde r}=0$ and satisfies $\dot p_r<0$, the coordinates $(r,\phi_0)$ are recorded. We used the values $m=0.5$ and $\mu_r=2$ for the numerical evaluation. The dimensionless $m$ parameter corresponds to the following values: $\eta\approx1\hspace{2px}ps$ and $K_\perp\approx 6.6\cdot10^{-22}J=4.1\hspace{2px}\text{meV}$. The field-like driving $\mu_r$ is achieved by using an external magnetic field that is similar to the in-plane anisotropy. If we assume that $\gamma\approx2.8\cdot 10^{11}\hspace{2px}\text{Hz}\hspace{2px}T^{-1}$, then this would imply $B\approx0.3\hspace{2px}T$.}
    \label{fig:PoincareSection}
\end{figure}

\noindent They describe a four-dimensional vector field, with trajectories that solve Eq. (\ref{Eq: Eq of motion vector notation}) without friction.
Fig. \ref{fig:PoincareSection} visualises 33 testing trajectories through a Poincar\'e section.
The coordinates $(\tilde r,\phi_0)$ of the trajectories are recorded when passing through the hyperplane $p_r=0$ and $\dot p_r<0$, and each trajectory has randomly chosen initial conditions.
The $\phi_0$ axes has been taken to allow only values between $0$ and $\pi$ radians, since the potential $V(\tilde r,\phi_0)$ is periodic in its second argument over $\pi$ radians.
While the Poincar\'e section provides a geometric snapshot of a dynamical system’s behavior, chaotic dynamics in classical Hamiltonian systems are characterized by sensitive dependence on initial conditions. 
A standard quantitative signature of chaos is a positive largest Lyapunov exponent (LLE), indicating that nearby trajectories diverge exponentially on average, leading to eventual loss of correlation between them. In contrast, regular motion yields an LLE of zero, implying stability.

\begin{figure}
    \centering
    \includegraphics[width=\linewidth]{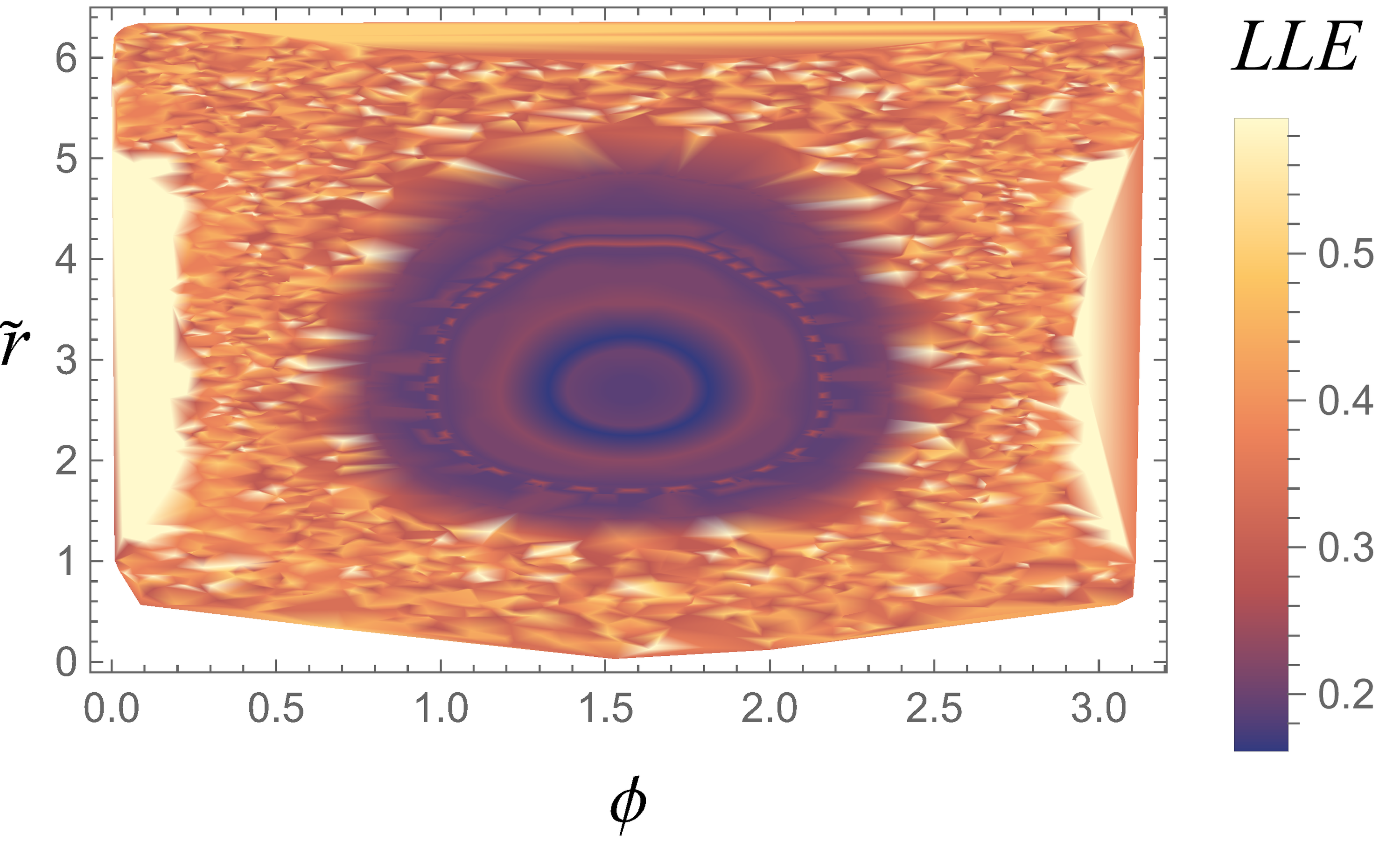}
    \caption{The correspondence between the largest Lyapunov exponent (LLE) and the Poincaré section. Parameters are identical to those used in Fig.~\ref{fig:PoincareSection}.}
    \label{fig:LLE}
\end{figure}

Therefore, we compute the LLE of the four-dimensional system given by Eq. (\ref{Eq: Eq of motion vector notation}).
We represent the set of first order differential equations by $\dot{x}=f(x),$ with state vector $x=(r,\phi,p_{r},p_{\phi})^{T}$.
This equation is non-linear and generally non-integrable for certain parameter choices, and it can exhibit chaos. To study the growth of small perturbations, we linearize the system about a reference trajectory. The Jacobian matrix $J(x)=Df(x)$ that describes the linearization of the vector field $f(x)$ reads

\begin{equation}
J(x)=\begin{bmatrix}0 & 0 & \tfrac{1}{m} & 0\\[4pt]
\tfrac{1}{m} & 0 & 0 & \tfrac{1}{m}\\[4pt]
-\tfrac{1}{m} & 0 & 0 & -\tfrac{1}{m}\\[4pt]
0 & -2\cos(2\phi) & 0 & 0
\end{bmatrix}.
\end{equation}

\noindent Let $x(t;x_{0})$ denote the solution trajectory starting from initial condition $x(0)=x_{0}$. To examine how an infinitesimal perturbation grows or decays along this trajectory, we define the deviation vector as $\delta x(t)=x(t;x_0+\varepsilon)-x(t;x_0)$, which for small $\varepsilon$ evolves according to the linearized dynamics $\dot{\delta x}(t)\approx J(x(t))\delta x(t)$. Neglecting $O(|\delta x|^{2})$ terms, we obtain the system

\begin{subequations}

\begin{align}
    \dot{\delta \tilde r} & =\tfrac{1}{m}\delta p_{r}, & \dot{\delta p_{r}} & =-\tfrac{1}{m}\delta r-\tfrac{1}{m}\delta p_{\phi}, \\
    \dot{\delta\phi} & =\tfrac{1}{m}\delta r+\tfrac{1}{m}\delta p_{\phi}, & \dot{\delta p_{\phi}} & =-2\cos\big(2\phi(t)\big)\delta\phi.
\end{align}    
\end{subequations}

\noindent The LLE, $\lambda_{1}$, measures mean exponential separation of nearby trajectories:

\begin{equation}
\|\delta x(t)\|\approx\|\delta x(0)\|e^{\lambda_{1}t}.    
\end{equation}

\noindent From the same trajectories that we generated the Poincaré section with, we compute the LLE using the Benettin--Wolf algorithm \cite{wolf1985determining,benettin1980lyapunov}. The LLE is positive throughout the section (Fig.~\ref{fig:LLE}), the central island shows reduced values, whereas the surrounding speckled region exhibits irregular spatial fluctuations of the positive exponent. The persistent positivity of the LLE indicates chaotic dynamics, i.e., sensitive dependence on initial conditions.

\section{Domain wall velocities}\label{sec:Domain wall velocities}

\begin{figure}
    \centering
    \includegraphics[width=\linewidth]{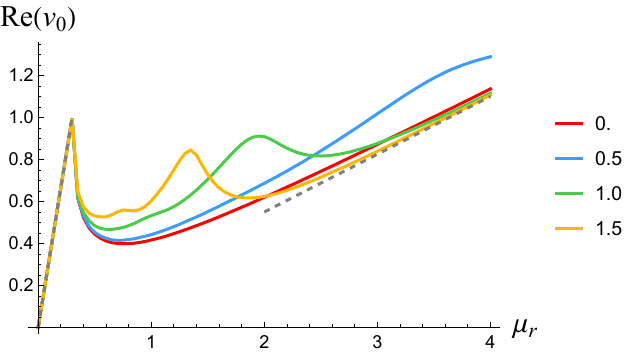}
    \caption{The DW velocity in function of field-like driving, $\mu_r$, for different values of $m=\{0,0.5,1.0,1.5\}$. The gray dashed lines are the analytically derived limits.}
    \label{fig:DWmass}
\end{figure}

\begin{figure}
    \centering
    \includegraphics[width=\linewidth]{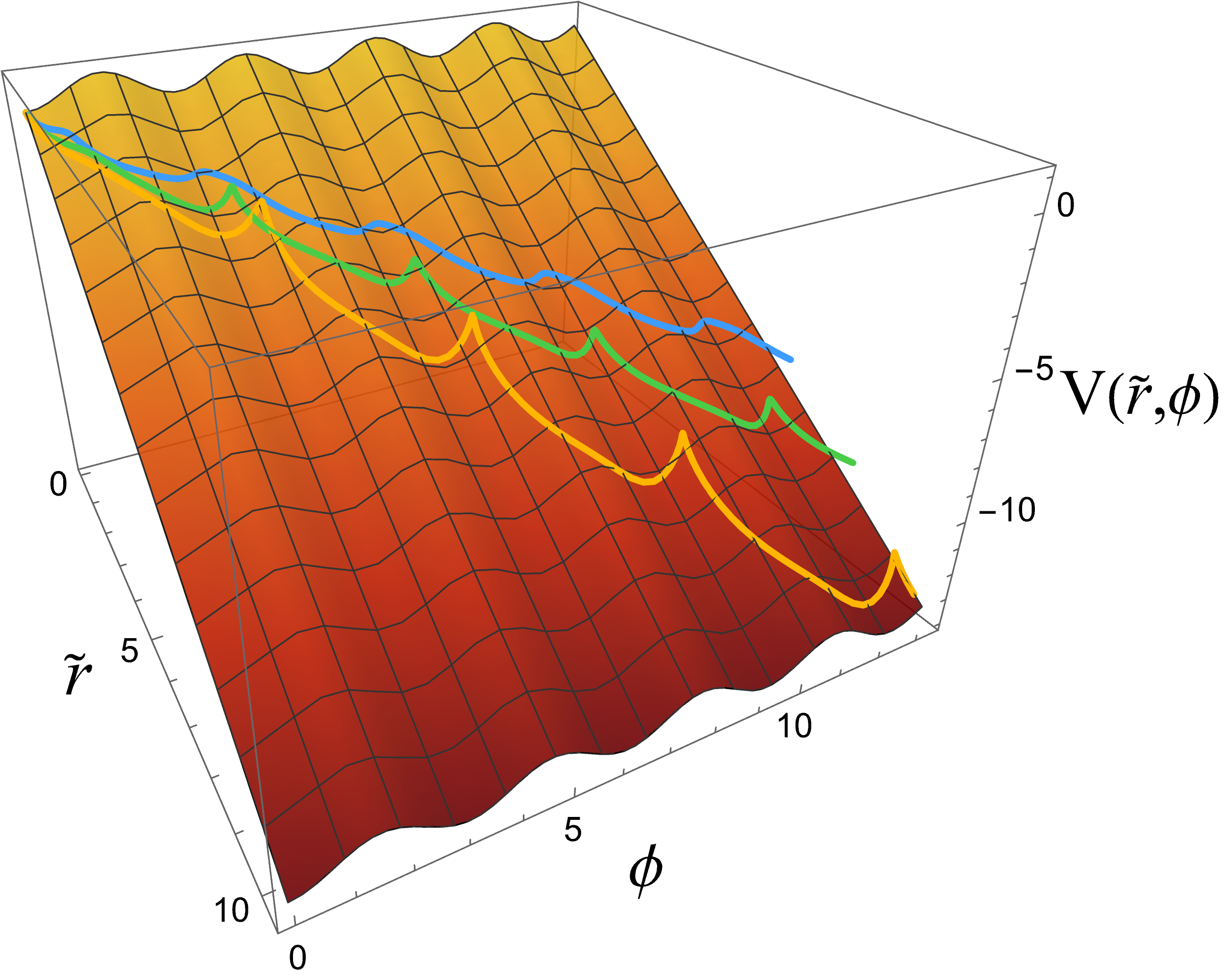}
    \caption{The trajectories on the potential landscape are given by the time evolution of the DW variables, $(\tilde r,\phi_0)$, at $\mu_r=1.35$. The colors match the normalised inertial parameter $m$ of Fig. \ref{fig:DWmass}.}
    \label{fig:DWtrajectory}
\end{figure}

In this section, we consider the effects of damping, which makes the DW motion regular.
We start by finding analytical bounds on the DW velocity and its behavior in the limit of small and large driving.
Thereafter, we numerically compute DW velocities for different types of driving, field-like and damping-like.
We show that for specific values of driving through field-like torques, the DW with inertial effects is faster than without.

We recast Eq. (\ref{Eq: Eq of motion vector notation}) in a complex equation,

\begin{equation}\label{eq:complexR}
    m\ddot{s} = (i-\alpha)\dot{s} +\mu-i\sin(2\text{Im}(s)),
\end{equation}

\noindent where $s=\tilde r+i\phi_0$ and $\mu=\mu_r-i\mu_\phi$, such that the field-like driving is the real part and the damping-like driving is the imaginary part.
We separate $s$ into a linear and an oscillating part, $s=v_0t+f(t)+c$, where the following holds

\begin{subequations}
    
\begin{align}
    v_0=&\frac{\alpha+i}{1+\alpha^2}(\mu-i\langle\sin\left(2\text{Im}(v_0t+f(t)+c)\right)\rangle), \\
    m\ddot{f} =& (i-\alpha)\dot{f}-i\sin\left(2\text{Im}(v_0t+f(t)+c)\right) \\
     &+i\langle\sin\left(2\text{Im}(v_0t+f(t)+c)\right)\rangle. \nonumber
\end{align}
\end{subequations}

\noindent The brackets denote time averaging, $\langle\dots\rangle=\int_0^T\dots\frac{dt}{T}$ for a large time interval $T$.
This ansatz effectively separates the linear movement from the (almost) periodic motion.
From this ansatz, we find that the velocity of the DW is given by,

\begin{align}\label{eq:velocity}
    \text{Re}(v_0)=&\frac{\alpha\mu_r+\mu_\phi+\langle\sin\left(2\text{Im}(v_0t+f(t)+c)\right)\rangle}{1+\alpha^2}.
\end{align}

\noindent The exact expression of $\text{Re}(v_0)$ requires the exact expression of $f(t)$.
However, finding this is outside the scope of our analysis.
Estimates can be made within the limits of large and small driving.
For large driving, $\alpha\mu_r+\mu_\phi\gg1$, we have that $\text{Re}(v_0)\approx(\alpha\mu_r+\mu_\phi)/(1+\alpha^2)$.
In contrast, the limit of small driving, $|\mu_r/\alpha-\mu_\phi|< 1$, considers the case that the driving force is insufficient to overcome the potential barrier that is set by $V(\tilde r,\phi_0)$ in the $\phi_0$ direction.
The critical driving that is necessary to overcome the potential barrier is known as Walker breakdown.
Before the Walker breakdown, we have that $f(t)=0$, which allows us to derive the relation $\langle\sin(2\phi_0)\rangle=\mu_r/\alpha-\mu_\phi$ from the definition of the brackets.
This expression solves Eq. (\ref{eq:velocity}), from which we find the velocity before the Walker breakdown, given by

\begin{equation}
    \text{Re}(v_0)=\frac{\mu_r}{\alpha}.
\end{equation}

\begin{figure}
    \centering
    \includegraphics[width=\linewidth]{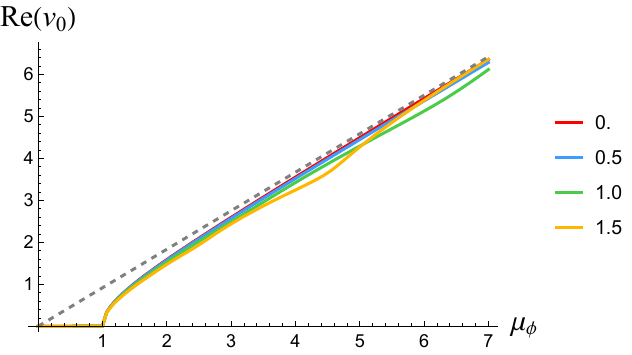}
    \caption{The DW velocity in function of damping-like driving, $\mu_\phi$, for different values of $m=\{0,0.5,1.0,1.5\}$. The gray dashed line is the analytically derived limit.}
    \label{fig:DWmassSOT}
\end{figure}

\noindent These analytical limits correspond well with the numerical evaluation of Eq. (\ref{eq:complexR}), as shown by the gray dashed lines in Figs. \ref{fig:DWmass} and \ref{fig:DWmassSOT}.
In these figures, we numerically compute the average velocity of a DW as a function of $\mu$ for different normalised inertial parameter $m$.
The dimensionless parameter is taken to be $\{0.5,1.0,1.5\}$ and $\alpha=0.3$.
These values for $m$ may correspond to an inertial relaxation time of $1\hspace{3px}ps$, which is close to experimental values of Ref. \cite{IRT:Neeraj_Awari_Kovalev_Polley_Zhou_Hagstrom_Arekapudi_Semisalova_Lenz_Green_Deinert_et_al._2021}, and an anisotropy $K_\perp\approx\{4.1,8.2,12.3\}\text{meV}$.
As the first case, we consider only field-like driving (Fig. \ref{fig:DWmass}).
Herein, we observe that the inertial DW velocity exhibits a distinct maximum, which is almost twice the velocity of the non-inertial DW at the same driving.
To investigate this further, we analyse the different trajectories of the DW parameters (Fig. \ref{fig:DWtrajectory}).
For the field-like driving $\mu_r=1.35$, we find an efficient trajectory from $m=1.5$, where its periodicity matches that of the potential.
While other trajectories, $m=0.5$ and $m=1.0$, do not coalesce with the potential.
The resonance of the cyclotronic motion with the periodic potential gives the inertial DW a greater velocity.
When considering damping-like driving, the potential does not allow for period coalescence.
Therefore, we do not observe maxima (Fig. \ref{fig:DWmassSOT}), but rather decreased velocities compared to the non-inertial case.

\begin{figure}
    \centering
    \includegraphics[width=\linewidth]{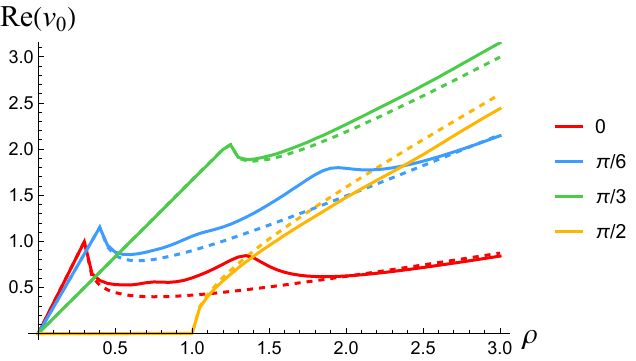}
    \caption{The DW velocity in function of total driving, $\mu=\rho e^{i\theta}$. This is calculated for different combinations of field-like and damping-like driving parametrized by $\theta$. The normalised inertial parameter is taken to be $m=1.5$. The dashed lines are the velocities of non-inertial DWs.}
    \label{fig:DWmassAngle}
\end{figure}

Lastly, we take $\mu=\rho e^{-i\theta}$ and vary $\rho$ for different $\theta$ (Fig. \ref{fig:DWmassAngle}).
This incorporates both field-like and damping-like driving, which is realistic for STT and SOT.
We find that field-like driving, including inertia, outperforms the non-inertial case, and this trend is seen for most combinations of field-like and damping-like driving.

\section{Spin inertia effect on the domain wall width}\label{sec:DWwidth}

We show that before the Walker breakdown, $|\mu_r/\alpha-\mu_\phi|\leq 1$, the DW width has an additional contribution from the inertia.
In this limit, the velocity of the DW is given by $\dot{\tilde{r}}=\mu_r/\alpha$ and $\dot\phi_0=0$.
We substitute these results into the action of Eq. \ref{eq:A} and take $\lambda$ arbitrary.
The effective action is given by,

\begin{align}
     \mathcal{A}_{reduced} =&\tilde N\int dt \Big\{\left(B+\sigma \tau\right)\frac{\mu_r t}{\alpha}+\hbar \phi_0(t)\frac{\mu_r}{\alpha}-\hbar v_s\phi_0\nonumber\\
     &+\frac{\hbar\eta}{2\alpha^2}\frac{\mu_r^2}{\lambda}-\frac{J_s}{\lambda}-\lambda K_z-\lambda\frac{K_\perp}{2}\sin^2(\phi_0)\Big\},
\end{align}

\noindent with $\tilde N=2 S/a^3$.
This reduced action is minimized when $\delta\mathcal{A}_{reduced}/\delta \phi_0=\delta\mathcal{A}_{reduced}/\delta \lambda=0$.
The minimization leads to the following equations,

\begin{subequations}
\begin{align}
    \frac{K_\perp}{2\hbar}\sin (2\phi_0)=&\frac{\mu_r-\alpha v_s}{\alpha\lambda},\\
    \lambda=&\sqrt{\frac{J_s-\hbar\eta\mu_r^2/2\alpha^2}{K_z+K_\perp \sin^2(\phi_0)}}.
\end{align}    
\end{subequations}

\noindent The last equality gives the modified DW width, which is decreased due to spin-inertia.

\section{Discussion and Conclusion}\label{sec:Discussion and Conclusion}

We have shown that incorporating spin inertia into the magnetization dynamics modifies the collective-coordinate dynamics of a ferromagnetic DW. In particular, the coupled variables $(r(t),\phi_{0}(t))$ acquire second-order time derivatives, which endow the DW with an effective mass set by the inertial parameter $\eta$. Written in vector form, the equations map onto the dynamics of a charged particle moving in a two-dimensional potential under an effective perpendicular magnetic field and damping. In the dissipationless limit, this mapping yields Hamiltonian dynamics and, for the parameter regimes explored here, a positive largest Lyapunov exponent, demonstrating sensitive dependence on initial conditions and hence chaotic motion. For finite damping, the same structure persists but damping regularizes the trajectories and suppresses the chaotic response, causing the long-time dynamics responsible to averages such as the mean DW velocity.

A main consequence of the inertial term is the emergence of resonant enhancements of the DW velocity under field-like driving. Numerically evaluating the dynamics shows pronounced maxima of the average velocity as a function of the real driving component $\mu_{r}$, with peak velocities approaching nearly twice those of the non-inertial case for the representative parameters studied (Fig.~$\ref{fig:DWmass}$). The study of the corresponding trajectories (Fig.~$\ref{fig:DWtrajectory}$) indicates that these maxima arise from a proportionality between the motion in $(\widetilde{r},\phi_{0})$ space and the periodic structure of the effective potential, which enables particularly efficient transport through phase space. In contrast, purely damping-like driving given by the imaginary component of $\mu$ does not permit such period matching and therefore does not generate analogous maxima. Instead, the inertial contribution tends to reduce the average velocity relative to the non-inertial case (Fig.~$\ref{fig:DWmassSOT}$). For mixed driving, as expected for STT and SOT, inertial DWs typically retain a velocity advantage over non-inertial DWs for most combinations of field-like and damping-like components (Fig.~$\ref{fig:DWmassAngle}$), with degradation primarily when damping-like torques dominate.

Spin inertia also modifies the domain-wall width. Below the Walker-breakdown threshold, we obtained an explicit expression for the inertial renormalization of the width and found a reduction compared to the standard result. This provides a complementary and potentially practical signature of inertia, in parameter regimes where velocity-based signatures are masked by pinning or sample-dependent dissipation, a systematic dependence of the wall width on driving could offer an additional route to infer inertial effects.

The present results are obtained within a minimal collective-coordinate model and therefore define a clear set of extensions. Disorder, pinning landscapes, thermal fluctuations, nonuniform wall deformations, and additional internal modes may alter the quantitative location and sharpness of velocity maxima, and they may further regularize dynamics that are chaotic in the idealized, dissipationless limit. Nevertheless, the mechanisms identified here are robust at the level of equations of motion: (i) the spin inertial term generates an effective DW mass; (ii) damping suppresses chaos and leads to regular long-time motion; and (iii) in the case of field-like driving, spin inertia enhances the average velocity in comparison to massless DW motion.

These features motivate micromagnetic simulations and experimental verifications in materials or heterostructures exhibiting relatively large inertial time. If such regimes can be realized, inertial DWs may provide an avenue for higher-speed operation at comparable drive strengths, which is relevant for DW based device concepts such as racetrack memory.

\begin{acknowledgments}
A.L.B and R.A.D. acknowledge financial support by the projects “Black holes on a chip” with project number OCENW.KLEIN.502 and “Fluid Spintronics” with project number VI.C.182.069.
Both are financed by the Dutch Research Council (NWO).
M.C. acknowledges funding from the Deutsche Furschungsgemeinschaft (DFG, German Research Foundation) under the FOR 5844: ChiPS, Project-ID 541503763.
P.-B.H. acknowledges financial support by the National Science Foundation (NSF) of Changsha City (Grant No. kq2208008) and the NSF of Hunan Province (Grant No. 2023JJ30116).
\end{acknowledgments}



\newpage

\nocite{*}

\bibliography{bib}

@article{Suhl1998,
  author={Suhl, H.},
  journal={IEEE Trans. Magn.}, 
  title={Theory of the magnetic damping constant}, 
  year={1998},
  volume={34},
  number={4},
  pages={1834-1838},
  doi={10.1109/20.706720}}

@article{wegrowe2000thermokinetic,
  title = {{Thermokinetic approach of the generalized Landau-Lifshitz-Gilbert equation with spin-polarized current}},
  author = {Wegrowe, J.-E.},
  journal = {Phys. Rev. B},
  volume = {62},
  issue = {2},
  pages = {1067--1074},
  numpages = {0},
  year = {2000},
  month = {Jul},
  publisher = {American Physical Society},
  doi = {10.1103/PhysRevB.62.1067},
  url = {https://link.aps.org/doi/10.1103/PhysRevB.62.1067}
}

@article{Ciornei2011,
  author =        {Ciornei, M.-C. and Rub\'{\i}, J. M. and
                   Wegrowe, J.-E.},
  journal =       {Phys. Rev. B},
  month =         {Jan},
  pages =         {020410},
  publisher =     {American Physical Society},
  title =         {Magnetization dynamics in the inertial regime:
                   Nutation predicted at short time scales},
  volume =        {83},
  year =          {2011},
  doi =           {10.1103/PhysRevB.83.020410},
  url =           {http://link.aps.org/doi/10.1103/PhysRevB.83.020410},
}

@article{Wegrowe2012,
  author =        {J.-E. Wegrowe and M.-C. Ciornei},
  journal =       {Am. J. Phys.},
  number =        {7},
  pages =         {607-611},
  title =         {Magnetization dynamics, gyromagnetic relation, and
                   inertial effects},
  volume =        {80},
  year =          {2012},
  doi =           {10.1119/1.4709188},
  url =           {http://dx.doi.org/10.1119/1.4709188},
}

@article{giordano2020derivation,
  title = {Derivation of magnetic inertial effects from the classical mechanics of a circular current loop},
  author = {Giordano, Stefano and D\'ejardin, Pierre-Michel},
  journal = {Phys. Rev. B},
  volume = {102},
  issue = {21},
  pages = {214406},
  numpages = {13},
  year = {2020},
  month = {Dec},
  publisher = {American Physical Society},
  doi = {10.1103/PhysRevB.102.214406},
  url = {https://link.aps.org/doi/10.1103/PhysRevB.102.214406}
}

@article{Fahnle2011,
  author =        {F\"ahnle, Manfred and Steiauf, Daniel and
                   Illg, Christian},
  journal =       {Phys. Rev. B},
  month =         {Nov},
  pages =         {172403},
  publisher =     {American Physical Society},
  title =         {{Generalized Gilbert equation including inertial
                   damping: Derivation from an extended breathing Fermi
                   surface model}},
  volume =        {84},
  year =          {2011},
  doi =           {10.1103/PhysRevB.84.172403},
  url =           {http://link.aps.org/doi/10.1103/PhysRevB.84.172403},
}

@article{Bhattacharjee2012,
  title = {Atomistic spin dynamic method with both damping and moment of inertia effects included from first principles},
  author = {Bhattacharjee, Satadeep and Nordstr\"om, Lars and Fransson, Jonas},
  journal = {Phys. Rev. Lett.},
  volume = {108},
  issue = {5},
  pages = {057204},
  numpages = {5},
  year = {2012},
  month = {Jan},
  publisher = {American Physical Society},
  doi = {10.1103/PhysRevLett.108.057204},
  url = {https://link.aps.org/doi/10.1103/PhysRevLett.108.057204}
}

@article{Mondal2017Nutation,
  title = {Relativistic theory of magnetic inertia in ultrafast spin dynamics},
  author = {Mondal, Ritwik and Berritta, Marco and Nandy, Ashis K. and Oppeneer, Peter M.},
  journal = {Phys. Rev. B},
  volume = {96},
  issue = {2},
  pages = {024425},
  numpages = {9},
  year = {2017},
  month = {Jul},
  publisher = {American Physical Society},
  doi = {10.1103/PhysRevB.96.024425},
  url = {https://link.aps.org/doi/10.1103/PhysRevB.96.024425}
}

@article{Mondal2018JPCM,
  author={Ritwik Mondal and Marco Berritta and Peter M Oppeneer},
  title={Generalisation of {G}ilbert damping and magnetic inertia parameter as a series of higher-order relativistic terms},
  journal={J. Phys.: Condens. Matter},
  volume={30},
  number={26},
  pages={265801},
  url={http://stacks.iop.org/0953-8984/30/i=26/a=265801},
  year={2018}
}

@article{SBhattacharjee,
  author = {Bhattacharjee, Satadeep and Nordstr\"{o}m, Lars and Fransson, Jonas},
  year = {2012},
  title = {Atomistic Spin Dynamic Method with both Damping and Moment of Inertia Effects Included from First Principles},
  journal = {Phys. Rev. Lett.},
  volume = {108},
  number = {5},
  pages = {057204},
  doi = {10.1103/physrevlett.108.057204},
  url = {http://dx.doi.org/10.1103/physrevlett.108.057204},
}

@article{RMondal_JPCM,
  author = {Mondal, Ritwik and Berritta, Marco and Oppeneer, Peter M},
  year = {2018},
  title = {Generalisation of Gilbert damping and magnetic inertia parameter as a series of higher-order relativistic terms},
  journal = {J. Phys.: Condens. Matter},
  volume = {30},
  number = {26},
  pages = {265801},
  doi = {10.1088/1361-648x/aac5a2},
  url = {http://dx.doi.org/10.1088/1361-648x/aac5a2},
}

@article{RMondal,
  author = {Mondal, Ritwik and R\'{o}zsa, Levente and Farle, Michael and Oppeneer, Peter M. and Nowak, Ulrich and Cherkasskii, Mikhail},
  year = {2023},
  title = {Inertial effects in ultrafast spin dynamics},
  journal = {J. Magn. Magn. Mater.},
  volume = {579},
  month = {August},
  pages = {170830},
  doi = {10.1016/j.jmmm.2023.170830},
  url = {http://dx.doi.org/10.1016/j.jmmm.2023.170830},
}

@article{MAQuarenta,
  author = {Quarenta, Mario Gaspar and Tharmalingam, Mithuss and Ludwig, Tim and Yuan, H. Y. and Karwacki, Lukasz and Verstraten, Robin C. and Duine, Rembert A.},
  year = {2024},
  title = {Bath-Induced Spin Inertia},
  journal = {Phys. Rev. Lett.},
  volume = {133},
  number = {13},
  pages = {136701},
  doi = {10.1103/physrevlett.133.136701},
  url = {http://dx.doi.org/10.1103/physrevlett.133.136701},
}

@article{JAnders,
  author = {Anders, J and Sait, C R J and Horsley, S A R},
  year = {2022},
  title = {Quantum Brownian motion for magnets},
  journal = {New J. Phys.},
  volume = {24},
  number = {3},
  pages = {033020},
  doi = {10.1088/1367-2630/ac4ef2},
  url = {http://dx.doi.org/10.1088/1367-2630/ac4ef2},
}

@article{TKikuchi,
  author = {Kikuchi, Toru and Tatara, Gen},
  year = {2015},
  title = {Spin dynamics with inertia in metallic ferromagnets},
  journal = {Phys. Rev. B},
  volume = {92},
  number = {18},
  pages = {184410},
  doi = {10.1103/physrevb.92.184410},
  url = {http://dx.doi.org/10.1103/physrevb.92.184410},
}

@article{EOlive_APL,
  author = {Olive, E. and Lansac, Y. and Wegrowe, J.-E.},
  year = {2012},
  title = {Beyond ferromagnetic resonance: The inertial regime of the magnetization},
  journal = {Appl. Phys. Lett.},
  volume = {100},
  number = {19},
  pages = {192407},
  doi = {10.1063/1.4712056},
  url = {http://dx.doi.org/10.1063/1.4712056},
}

@article{EOlive_JAP,
  author = {Olive, E. and Lansac, Y. and Meyer, M. and Hayoun, M. and Wegrowe, J.-E.},
  year = {2015},
  title = {Deviation from the Landau-Lifshitz-Gilbert equation in the inertial regime of the magnetization},
  journal = {J. Appl. Phys.},
  volume = {117},
  number = {21},
  pages = {213904},
  doi = {10.1063/1.4921908},
  url = {http://dx.doi.org/10.1063/1.4921908},
}

@article{VUnikandanunni,
  author = {Unikandanunni, Vivek and Medapalli, Rajasekhar and Asa, Marco and Albisetti, Edoardo and Petti, Daniela and Bertacco, Riccardo and Fullerton, Eric E. and Bonetti, Stefano},
  year = {2022},
  title = {Inertial spin dynamics in epitaxial cobalt films},
  journal = {Phys. Rev. Lett.},
  volume = {129},
  number = {23},
  pages = {237201},
  doi = {10.1103/physrevlett.129.237201},
  url = {http://dx.doi.org/10.1103/physrevlett.129.237201},
}

@article{KNeeraj,
  author = {Neeraj, Kumar and Awari, Nilesh and Kovalev, Sergey and Polley, Debanjan and Zhou Hagstr\"{o}m, Nanna and Arekapudi, Sri Sai Phani Kanth and Semisalova, Anna and Lenz, Kilian and Green, Bertram and Deinert, Jan-Christoph and Ilyakov, Igor and Chen, Min and Bawatna, Mohammed and Scalera, Valentino and d’Aquino, Massimiliano and Serpico, Claudio and Hellwig, Olav and Wegrowe, Jean-Eric and Gensch, Michael and Bonetti, Stefano},
  year = {2021},
  title = {Inertial spin dynamics in ferromagnets},
  journal = {Nature Physics},
  volume = {17},
  number = {2},
  pages = {245-250},
  doi = {10.1038/s41567-020-01040-y},
  url = {http://dx.doi.org/10.1038/s41567-020-01040-y},
}

@article{YLi,
  author = {Li, Y. and Barra, A.-L. and Auffret, S. and Ebels, U. and Bailey, W. E.},
  title = {Inertial terms to magnetization dynamics in ferromagnetic thin films},
  journal = {Phys. Rev. B},
  volume = {92},
  issue = {14},
  pages = {140413},
  numpages = {5},
  year = {2015},
  month = {Oct},
  publisher = {American Physical Society},
  doi = {10.1103/PhysRevB.92.140413},
  url = {https://link.aps.org/doi/10.1103/PhysRevB.92.140413}
}

@article{ADe,
  author = {De, Anulekha and Schlegel, Julius and Lentfert, Akira and Scheuer, Laura and Stadtm\"{u}ller, Benjamin and Pirro, Philipp and von Freymann, Georg and Nowak, Ulrich and Aeschlimann, Martin},
  year = {2025},
  title = {Magnetic nutation: Transient separation of magnetization from its angular momentum},
  journal = {Phys. Rev. B},
  volume = {111},
  number = {1},
  pages = {014432},
  doi = {10.1103/physrevb.111.014432},
  url = {http://dx.doi.org/10.1103/physrevb	.111.014432},
}

@article{SVTitov,
  author = {Titov, S. V. and Kalmykov, Yu. P. and Kazarinov, K. D. and Cherkasskii, M. A. and Titov, A. S.},
  year = {2023},
  title = {Inertial Magnetization Dynamics in Ferromagnetic Nanoparticles Near Saturation},
  journal = {J. Commun. Technol. Electron.},
  volume = {68},
  number = {5},
  pages = {559-565},
  doi = {10.1134/s1064226923050169},
  url = {http://dx.doi.org/10.1134/s1064226923050169},
}

@article{MCherkasskii_PRB102,
  author = {Cherkasskii, Mikhail and Farle, Michael and Semisalova, Anna},
  year = {2020},
  title = {Nutation resonance in ferromagnets},
  journal = {Phys. Rev. B},
  volume = {102},
  number = {18},
  pages = {184432},
  doi = {10.1103/physrevb.102.184432},
  url = {http://dx.doi.org/10.1103/physrevb.102.184432},
}

@article{MCherkasskii_PRB106,
  author = {Cherkasskii, Mikhail and Barsukov, Igor and Mondal, Ritwik and Farle, Michael and Semisalova, Anna},
  year = {2022},
  title = {Theory of inertial spin dynamics in anisotropic ferromagnets},
  journal = {Phys. Rev. B},
  volume = {106},
  number = {5},
  pages = {054428},
  doi = {10.1103/physrevb.106.054428},
  url = {http://dx.doi.org/10.1103/physrevb.106.054428},
}

@article{RRodriguez,
  author = {Rodriguez, Rodolfo and Cherkasskii, Mikhail and Jiang, Rundong and Mondal, Ritwik and Etesamirad, Arezoo and Tossounian, Allison and Ivanov, Boris A. and Barsukov, Igor},
  year = {2024},
  title = {Spin Inertia and Auto-Oscillations in Ferromagnets},
  journal = {Phys. Rev. Lett.},
  volume = {132},
  number = {24},
  pages = {246701},
  doi = {10.1103/physrevlett.132.246701},
  url = {http://dx.doi.org/10.1103/physrevlett.132.246701},
}

@article{PBHe_PRB108,
  author = {He, Peng-Bin},
  year = {2023},
  title = {Large-amplitude and widely tunable self-oscillations enabled by the inertial effect in uniaxial antiferromagnets driven by spin-orbit torques},
  journal = {Phys. Rev. B},
  volume = {108},
  number = {18},
  pages = {184418},
  doi = {10.1103/physrevb.108.184418},
  url = {http://dx.doi.org/10.1103/physrevb.108.184418},
}

@article{PBHe_PRB110,
  author = {P.-B. He},
  year = {2024},
  title = {Influence of the magnetic inertia on the self-oscillation in spin-orbit torque-driven tripartite antiferromagnets with a $120^\circ$ rotation symmetry},
  journal = {Phys. Rev. B},
  volume = {110},
  pages = {064411},
}

@article{benettin1980lyapunov,
  title={Lyapunov characteristic exponents for smooth dynamical systems and for Hamiltonian systems; a method for computing all of them. Part 1: Theory},
  author={Benettin, Giancarlo and Galgani, Luigi and Giorgilli, Antonio and Strelcyn, Jean-Marie},
  journal={Meccanica},
  volume={15},
  number={1},
  pages={9--20},
  year={1980},
  publisher={Springer}
}

@article{wolf1985determining,
  title={Determining Lyapunov exponents from a time series},
  author={Wolf, Alan and Swift, Jack B and Swinney, Harry L and Vastano, John A},
  journal={Physica D: nonlinear phenomena},
  volume={16},
  number={3},
  pages={285--317},
  year={1985},
  publisher={Elsevier}
}

@article{Shibata_Tatara_Kohno_2011, title={A brief review of field- and current-driven domain-wall motion}, volume={44}, url={https://iopscience.iop.org/article/10.1088/0022-3727/44/38/384004}, DOI={10.1088/0022-3727/44/38/384004}, abstractNote={A brief review of field- and recently developed current-driven domain-wall motion in a ferromagnetic nanowire is presented from a theoretical point of view. In the first part, the wall motion driven by an external magnetic field is studied on the basis of the Landau–Lifshitz–Gilbert equation and the collective coordinate method. The domain wall is treated as planar and rigid, called a one-dimensional model, and the wall motion is described by the relevant collective coordinates, centre position                X                and the polarization angle ϕ                0                of the wall. We also consider the interaction between the collective coordinates and spin waves excited around the wall and provide applicable criteria for the collective coordinate method in the domain-wall system. In the second part, we devote ourselves to studying the effect of conduction electrons on the domain-wall dynamics in a ferromagnetic metal. Microscopic calculations of the spin-transfer torque, dissipative spin torque (β-term), non-adiabatic force and Gilbert damping are presented on the basis of the linear response theory and its extension. In the third part, the current-driven domain-wall motion described by the collective coordinates is studied. The effect of external pinning is also examined. There are several depinning mechanisms and threshold currents in different pinning regimes.}, number={38}, journal={Journal of Physics D: Applied Physics}, author={Shibata, Junya and Tatara, Gen and Kohno, Hiroshi}, year={2011}, month={sept}, pages={384004} }

@article{Tatara_Kohno_Shibata_2008, title={Microscopic approach to current-driven domain wall dynamics}, volume={468}, url={https://linkinghub.elsevier.com/retrieve/pii/S0370157308002597}, DOI={10.1016/j.physrep.2008.07.003}, number={6}, journal={Physics Reports}, author={Tatara, Gen and Kohno, Hiroshi and Shibata, Junya}, year={2008}, pages={213–301} }

@article{Berger_1978, title={Low-field magnetoresistance and domain drag in ferromagnets}, volume={49}, url={https://pubs.aip.org/jap/article/49/3/2156/169108/Low-field-magnetoresistance-and-domain-drag-in}, DOI={10.1063/1.324716}, abstractNote={Despite common misconceptions, domain walls are too thick to ’’scatter’’ electrons appreciably. However, electrons crossing a wall apply a torque to it, which tends to cant the wall spins. This could be used to measure the conduction electron spin polarization. Most of the low-field resistive anomalies observed in pure Fe, Ni and Co at low temperature are caused by the Lorentz force associated with the internal field B=Ms present inside each domain. The existence of low-resistivity paths extending over many domains accounts for still unexplained magnetoresistance data in iron whiskers. In uniaxial materials, a d.c. eddy-current loop caused by the Hall effect runs around each wall. The field Hz generated by these loops tends to ’’drag’’ the whole domain structure in the direction of the carrier drift velocity. Also, the Joule dissipation of the eddy currents manifests itself as an excess Ohmic resistance. As predicted, this excess resistance decreases as the square of the field, in amorphous Gd25Co75 films, in MnBi films, and in pure bulk cobalt, when the walls are removed by an external field. The excess resistance can also be changed by reorienting the walls.}, number={3}, journal={Journal of Applied Physics}, author={Berger, L.}, year={1978}, month=mar, pages={2156–2161} }

@article{Schryer_Walker_1974, title={The motion of 180° domain walls in uniform dc magnetic fields}, volume={45}, url={https://pubs.aip.org/jap/article/45/12/5406/168816/The-motion-of-180-domain-walls-in-uniform-dc}, DOI={10.1063/1.1663252}, abstractNote={The equations of motion of a 180° domain wall in an infinite uniaxially anisotropic medium which is exposed to an instantaneously applied uniform dc magnetic field H0 have been integrated numerically. Below the critical field Hc =2παM0 (α is the Gilbert loss parameter and M0 the saturation magnetization), where a steady-state solution is known to exist, it is shown that the wall motion tends smoothly to this solution. Above Hc, the magnetization precesses about the field and a periodic component appears in the forward motion of the wall. Analytic solutions for the wall motion have been found based upon approximations suggested by the computed behavior; these reproduce the computer results very accurately.}, number={12}, journal={Journal of Applied Physics}, author={Schryer, N. L. and Walker, L. R.}, year={1974}, month={dec}, pages={5406–5421} }

@article{RT:Parkin_Hayashi_Thomas_2008, title={Magnetic Domain-Wall Racetrack Memory}, volume={320}, url={https://www.science.org/doi/10.1126/science.1145799}, DOI={10.1126/science.1145799}, abstractNote={Recent developments in the controlled movement of domain walls in magnetic nanowires by short pulses of spin-polarized current give promise of a nonvolatile memory device with the high performance and reliability of conventional solid-state memory but at the low cost of conventional magnetic disk drive storage. The racetrack memory described in this review comprises an array of magnetic nanowires arranged horizontally or vertically on a silicon chip. Individual spintronic reading and writing nanodevices are used to modify or read a train of ∼10 to 100 domain walls, which store a series of data bits in each nanowire. This racetrack memory is an example of the move toward innately three-dimensional microelectronic devices.}, number={5873}, journal={Science}, author={Parkin, Stuart S. P. and Hayashi, Masamitsu and Thomas, Luc}, year={2008}, month=apr, pages={190–194} }

@article{RT:Parkin_Yang_2015, title={Memory on the racetrack}, volume={10}, url={https://www.nature.com/articles/nnano.2015.41}, DOI={10.1038/nnano.2015.41}, number={3}, journal={Nature Nanotechnology}, author={Parkin, Stuart and Yang, See-Hun}, year={2015}, pages={195–198} }

@article{RT:Blasing_Khan_Filippou_Garg_Hameed_Castrillon_Parkin_2020, title={Magnetic Racetrack Memory: From Physics to the Cusp of Applications Within a Decade}, volume={108}, ISSN={1558-2256}, url={https://ieeexplore.ieee.org/document/9045991/}, DOI={10.1109/JPROC.2020.2975719}, abstractNote={Racetrack memory (RTM) is a novel spintronic memory-storage technology that has the potential to overcome fundamental constraints of existing memory and storage devices. It is unique in that its core differentiating feature is the movement of data, which is composed of magnetic domain walls (DWs), by short current pulses. This enables more data to be stored per unit area compared to any other current technologies. On the one hand, RTM has the potential for mass data storage with unlimited endurance using considerably less energy than today’s technologies. On the other hand, RTM promises an ultrafast nonvolatile memory competitive with static random access memory (SRAM) but with a much smaller footprint. During the last decade, the discovery of novel physical mechanisms to operate RTM has led to a major enhancement in the efficiency with which nanoscopic, chiral DWs can be manipulated. New materials and artificially atomically engineered thin-film structures have been found to increase the speed and lower the threshold current with which the data bits can be manipulated. With these recent developments, RTM has attracted the attention of the computer architecture community that has evaluated the use of RTM at various levels in the memory stack. Recent studies advocate RTM as a promising compromise between, on the one hand, power-hungry, volatile memories and, on the other hand, slow, nonvolatile storage. By optimizing the memory subsystem, significant performance improvements can be achieved, enabling a new era of cache, graphical processing units, and high capacity memory devices. In this article, we provide an overview of the major developments of RTM technology from both the physics and computer architecture perspectives over the past decade. We identify the remaining challenges and give an outlook on its future.}, number={8}, journal={Proceedings of the IEEE}, author={Bläsing, Robin and Khan, Asif Ali and Filippou, Panagiotis Ch. and Garg, Chirag and Hameed, Fazal and Castrillon, Jeronimo and Parkin, Stuart S. P.}, year={2020}, month={aug}, pages={1303–1321} }

@article{Mass:Saitoh_Miyajima_Yamaoka_Tatara_2004, title={Current-induced resonance and mass determination of a single magnetic domain wall}, volume={432}, url={https://www.nature.com/articles/nature03009}, DOI={10.1038/nature03009}, number={7014}, journal={Nature}, author={Saitoh, Eiji and Miyajima, Hideki and Yamaoka, Takehiro and Tatara, Gen}, year={2004}, pages={203–206} }

@article{Mass:Hurst_Galitski_Heikkila_2020, title={Electron-induced massive dynamics of magnetic domain walls}, volume={101}, url={https://link.aps.org/doi/10.1103/PhysRevB.101.054407}, DOI={10.1103/PhysRevB.101.054407}, number={5}, journal={Physical Review B}, author={Hurst, Hilary M. and Galitski, Victor and Heikkilä, Tero T.}, year={2020}, month={feb}, pages={054407} }

@article{Mass:doring1948tragheit,
  title={{\"U}ber die tr{\"a}gheit der w{\"a}nde zwischen wei{\ss}schen bezirken},
  author={D{\"o}ring, Werner},
  journal={Zeitschrift f{\"u}r Naturforschung A},
  volume={3},
  number={7},
  pages={373--379},
  year={1948},
  publisher={Verlag der Zeitschrift f{\"u}r Naturforschung}
}

@article{Mass:Janak_1964, title={Quantum Theory of Domain-Wall Motion}, volume={134}, rights={http://link.aps.org/licenses/aps-default-license}, ISSN={0031-899X}, url={https://link.aps.org/doi/10.1103/PhysRev.134.A411}, DOI={10.1103/PhysRev.134.A411}, number={2A}, journal={Physical Review}, author={Janak, James F.}, year={1964}, month={apr}, pages={A411–A422} }

@article{Chaos:Wagenhuber_Geisel_Niebauer_Obermair_1992, title={Chaos and anomalous diffusion of ballistic electrons in lateral surface superlattices}, volume={45}, url={https://link.aps.org/doi/10.1103/PhysRevB.45.4372}, DOI={10.1103/PhysRevB.45.4372}, number={8}, journal={Physical Review B}, author={Wagenhuber, J. and Geisel, T. and Niebauer, P. and Obermair, G.}, year={1992}, month={feb}, pages={4372–4383} }

@article{Chaos:Petschel_Geisel_1993, title={Bloch electrons in magnetic fields: Classical chaos and Hofstadter’s butterfly}, volume={71}, url={https://link.aps.org/doi/10.1103/PhysRevLett.71.239}, DOI={10.1103/PhysRevLett.71.239}, number={2}, journal={Physical Review Letters}, author={Petschel, G. and Geisel, T.}, year={1993}, month={july}, pages={239–242} }

@article{HofstadterButterfly,
  title = {Energy levels and wave functions of Bloch electrons in rational and irrational magnetic fields},
  author = {Hofstadter, Douglas R.},
  journal = {Phys. Rev. B},
  volume = {14},
  issue = {6},
  pages = {2239--2249},
  numpages = {0},
  year = {1976},
  month = {Sep},
  publisher = {American Physical Society},
  doi = {10.1103/PhysRevB.14.2239},
  url = {https://link.aps.org/doi/10.1103/PhysRevB.14.2239}
}

@article{STTandSOT:Seo_Kim_Ryu_Lee_Lee_2012, title={Current-induced motion of a transverse magnetic domain wall in the presence of spin Hall effect}, volume={101}, ISSN={0003-6951, 1077-3118}, url={https://pubs.aip.org/apl/article/101/2/022405/128156/Current-induced-motion-of-a-transverse-magnetic}, DOI={10.1063/1.4733674}, abstractNote={We theoretically study current-induced dynamics of a transverse magnetic domain wall in bi-layer nanowires consisting of a ferromagnetic layer on top of a nonmagnetic layer with strong spin-orbit coupling. Domain wall dynamics is characterized by two threshold current densities, JthWB and JthREV, where JthWB is a threshold for the chirality switching of the domain wall and JthREV is another threshold for the reversed domain wall motion caused by spin Hall effect. Domain walls with a certain chirality may move opposite to the electron-flow direction with high speed in the current range JthREV&lt;J&lt;JthWB for the system designed to satisfy the conditions JthWB&gt;JthREV and α&gt;β, where α is the Gilbert damping constant and β is the nonadiabaticity of spin torque. Micromagnetic simulations confirm the validity of analytical results.}, number={2}, journal={Applied Physics Letters}, author={Seo, Soo-Man and Kim, Kyoung-Whan and Ryu, Jisu and Lee, Hyun-Woo and Lee, Kyung-Jin}, year={2012}, month={july}, pages={022405} }

@article{DWChaos:Pivano_Dolocan_2016, title={Chaotic dynamics of magnetic domain walls in nanowires}, volume={93}, rights={http://link.aps.org/licenses/aps-default-license}, ISSN={2469-9950, 2469-9969}, url={https://link.aps.org/doi/10.1103/PhysRevB.93.144410}, DOI={10.1103/PhysRevB.93.144410}, number={14}, journal={Physical Review B}, author={Pivano, A. and Dolocan, V. O.}, year={2016}, month={apr}, pages={144410}}

@article{DWChaos:Suhl_Zhang_1987, title={Chaotic motion of domain walls in soft magnetic materials}, volume={61}, ISSN={0021-8979, 1089-7550}, url={https://pubs.aip.org/jap/article/61/8/4216/175065/Chaotic-motion-of-domain-walls-in-soft-magnetic}, DOI={10.1063/1.338479}, abstractNote={We have studied numerically one-dimensional rf driven motion of domain structures. The equation of motion solved has Landau–Lifshitz damping and includes all the basic phenomenological magnetic interactions: demagnetizing field, anisotropy field, and exchange field. We have found that for a large range of parameters, the spatial average of the magnetization is chaotic in time, and the spatial pattern at fixed time itself is likewise chaotic. The power spectrum of the chaotic time series has a 1/f shape. The phase boundary between chaotic and nonchaotic motion is described, and a limited analytical insight into this problem is discussed.}, number={8}, journal={Journal of Applied Physics}, author={Suhl, H. and Zhang, X. Y.}, year={1987}, month={apr}, pages={4216–4218}}

@article{DWChaos:Hermann_Nguenang_2013, title={Chaos Appearance during Domain Wall Motion under Electronic Transfer in Nanomagnets}, volume={03}, rights={http://creativecommons.org/licenses/by/4.0/}, ISSN={2160-6919, 2160-6927}, url={http://www.scirp.org/journal/doi.aspx?DOI=10.4236/wjcmp.2013.33022}, DOI={10.4236/wjcmp.2013.33022}, number={03}, journal={World Journal of Condensed Matter Physics}, author={Hermann, Donfack Gildas and Nguenang, Jean-Pierre}, year={2013}, pages={136–143} }

@article{DWChaos:Matsushita_Sasaki_Chawanya_2012, title={Chaos in AC-Driven Motion of Confined Magnetic Domain Wall}, volume={81}, ISSN={0031-9015, 1347-4073}, url={https://journals.jps.jp/doi/10.1143/JPSJ.81.063801}, DOI={10.1143/JPSJ.81.063801}, number={6}, journal={Journal of the Physical Society of Japan}, author={Matsushita, Katsuyoshi and Sasaki, Munetaka and Chawanya, Tsuyoshi}, year={2012}, month={june}, pages={063801}}

@article{DWChoas:Kosinski_Sukiennicki_1992, title={Chaotic motion of a domain wall in the time dependent drive fields}, volume={104–107}, rights={https://www.elsevier.com/tdm/userlicense/1.0/}, ISSN={03048853}, url={https://linkinghub.elsevier.com/retrieve/pii/030488539290820E}, DOI={10.1016/0304-8853(92)90820-E}, journal={Journal of Magnetism and Magnetic Materials}, author={Kosiński, R.A. and Sukiennicki, A.}, year={1992}, month={feb}, pages={331–332}}

@article{DWChaos:Okuno_Sugitani_Hirata_1995, title={Bifurcation diagram and power spectrum of chaotic domain-wall motion}, volume={140–144}, rights={https://www.elsevier.com/tdm/userlicense/1.0/}, ISSN={03048853}, url={https://linkinghub.elsevier.com/retrieve/pii/0304885394009279}, DOI={10.1016/0304-8853(94)00927-9}, journal={Journal of Magnetism and Magnetic Materials}, author={Okuno, H. and Sugitani, Y. and Hirata, K.}, year={1995}, month={feb}, pages={1879–1880}}

@article{DWChaos:Okuno_Hirata_Sakata_1995, title={Chaos of magnetic domain wall motion: Lyapunov exponent and controlling}, volume={31}, rights={https://ieeexplore.ieee.org/Xplorehelp/downloads/license-information/IEEE.html}, ISSN={00189464}, url={http://ieeexplore.ieee.org/document/490383/}, DOI={10.1109/20.490383}, number={6}, journal={IEEE Transactions on Magnetics}, author={Okuno, H. and Hirata, K. and Sakata, T.}, year={1995}, month={nov}, pages={3364–3366} }

@article{Yan2010,
  title = {Beating the Walker Limit with Massless Domain Walls in Cylindrical Nanowires},
  author = {Yan, Ming and K\'akay, Attila and Gliga, Sebastian and Hertel, Riccardo},
  journal = {Phys. Rev. Lett.},
  volume = {104},
  issue = {5},
  pages = {057201},
  numpages = {4},
  year = {2010},
  month = {Feb},
  publisher = {American Physical Society},
  doi = {10.1103/PhysRevLett.104.057201},
  url = {https://link.aps.org/doi/10.1103/PhysRevLett.104.057201}
}

@article{Yan_Andreas_Kakay_Garcia-Sanchez_Hertel_2011, title={Fast domain wall dynamics in magnetic nanotubes: Suppression of Walker breakdown and Cherenkov-like spin wave emission}, volume={99}, ISSN={0003-6951}, url={https://doi.org/10.1063/1.3643037}, DOI={10.1063/1.3643037}, abstractNote={We report on a micromagnetic study on domain wall (DW) propagation in ferromagnetic nanotubes. It is found that DWs in a tubular geometry are much more robust than ones in flat strips. This is explained by topological considerations. Our simulations show that the Walker breakdown of the DW can be completely suppressed. Constant DW velocities above 1000 m/s are achieved by small fields. A different velocity barrier of the DW propagation is encountered, which significantly reduces the DW mobility. This effect occurs as the DW reaches the phase velocity of spin waves (SWs), thereby triggering a Cherenkov-like emission of SWs.}, number={12}, journal={Applied Physics Letters}, author={Yan, Ming and Andreas, Christian and Kákay, Attila and García-Sánchez, Felipe and Hertel, Riccardo}, year={2011}, month={sept}, pages={122505} }

@article{Vogel2012,
  title = {Direct Observation of Massless Domain Wall Dynamics in Nanostripes with Perpendicular Magnetic Anisotropy},
  author = {Vogel, J. and Bonfim, M. and Rougemaille, N. and Boulle, O. and Miron, I. M. and Auffret, S. and Rodmacq, B. and Gaudin, G. and Cezar, J. C. and Sirotti, F. and Pizzini, S.},
  journal = {Phys. Rev. Lett.},
  volume = {108},
  issue = {24},
  pages = {247202},
  numpages = {5},
  year = {2012},
  month = {Jun},
  publisher = {American Physical Society},
  doi = {10.1103/PhysRevLett.108.247202},
  url = {https://link.aps.org/doi/10.1103/PhysRevLett.108.247202}
}

@article{Ciornei_Rubi_Wegrowe_2011, title={Magnetization dynamics in the inertial regime: Nutation predicted at short time scales}, volume={83}, rights={http://link.aps.org/licenses/aps-default-license}, ISSN={1098-0121, 1550-235X}, url={https://link.aps.org/doi/10.1103/PhysRevB.83.020410}, DOI={10.1103/PhysRevB.83.020410}, number={2}, journal={Physical Review B}, author={Ciornei, M.-C. and Rubí, J. M. and Wegrowe, J.-E.}, year={2011}, month={jan}, pages={020410}}

@article{IRT:Neeraj_Awari_Kovalev_Polley_Zhou_Hagstrom_Arekapudi_Semisalova_Lenz_Green_Deinert_et_al._2021, title={Inertial spin dynamics in ferromagnets}, volume={17}, ISSN={1745-2473, 1745-2481}, url={https://www.nature.com/articles/s41567-020-01040-y}, DOI={10.1038/s41567-020-01040-y}, number={2}, journal={Nature Physics}, author={Neeraj, Kumar and Awari, Nilesh and Kovalev, Sergey and Polley, Debanjan and Zhou Hagström, Nanna and Arekapudi, Sri Sai Phani Kanth and Semisalova, Anna and Lenz, Kilian and Green, Bertram and Deinert, Jan-Christoph and Ilyakov, Igor and Chen, Min and Bawatna, Mohammed and Scalera, Valentino and d’Aquino, Massimiliano and Serpico, Claudio and Hellwig, Olav and Wegrowe, Jean-Eric and Gensch, Michael and Bonetti, Stefano}, year={2021}, month={feb}, pages={245–250}}

@article{Ando_2008, title={Electric Manipulation of Spin Relaxation Using the Spin Hall Effect}, volume={101}, DOI={10.1103/PhysRevLett.101.036601}, number={3}, journal={Physical Review Letters}, author={Ando, K.}, year={2008} }

\end{document}